# Scalable fabrication of bijel films via continuous flow slit-coating

*Henrik Siegel[1,‡], Mariska de Ruiter[1,‡], Cos M. Hesseling[1], Georgios Athanasiou[1], Martin F. Haase[1]\**

[1] Van't Hoff Laboratory of Physical and Colloid Chemistry, Department of Chemistry, Debye Institute for Nanomaterials Science, Utrecht University, Utrecht, The Netherlands
‡ Both authors contributed equally
\* Correspondence: m.f.haase@uu.nl

**Abstract**

Nanocomposite membranes are an emerging filtration material in the production of clean water. Recently, the fabrication of porous nanocomposite membranes via solvent transfer induced phase separation (STrIPS) was introduced. During STrIPS, the phase separation of two immiscible liquids is arrested by the attachment of nanoparticles at the liquid-liquid interface, generating a porous particle-stabilized membrane template. STrIPS nanocomposite membranes, however, have so far only been produced as hollow fibers. To overcome this, we designed a roll-to-roll (R2R)-process for the continuous fabrication of flat-sheet nanocomposite membranes via STrIPS. We produce the STrIPS membrane by printing the membrane precursor on a carrier substrate. This allows for the continuous collection of STrIPS membranes with control over the membrane dimensions. Contact angle measurements elucidate the wetting dynamics of the STrIPS membrane on the substrate. Furthermore, we demonstrate control over the membrane pore structure by varying the precusor liquid composition and by changing the particle modification and loading in the membrane. With this, the R2R-approach may stimulate further advancements in the fabrication of flat-sheet nanocomposite membranes via STrIPS.

# 1. Introduction

The increasing scarcity and contamination of natural freshwater resources is stressing water supply systems worldwide [1]–[4]. To secure the demand for water with high quality standards, membrane technologies have been broadly implemented in water treatment processes [5]–[7]. Membranes separate dissolved or dispersed contaminants from water based on size or charge exclusion [8]. In view of the complexity and variety of water pollutants, nanocomposite membranes have emerged as a promising candidate in the development of new functional materials for water filtration [9]–[11].

Nanocomposite membranes are obtained by the incorporation of nano-entities such as nanoparticles into the membrane matrix [12]. The modification with nanoparticles equips the membrane with additional capabilities, for example enhanced anti-fouling properties or a high resistance to chemical cleaning procedures [13]. Furthermore, the addition of nanoparticles allows for control over the membrane porosity and membrane permeability [14]. An important prerequisite for regulating these membrane properties is the homogeneous dispersion of the nanoparticles in the membrane precursor as well as the uniform distribution of the nanoparticles in the resulting membrane matrix [15], [16].

A simple and recent approach addressing these requirements is the fabrication of nanocomposite membranes via solvent transfer induced phase separation (STrIPS) [17], [18]. During STrIPS, a homogeneous liquid mixture composed of oil, water, and a solvent is demixed by the controlled extraction of the solvent [19]. The demixing process is stabilized by the assembly of nanoparticles at the emerging liquid-liquid interface, driven by the reduction of the interfacial energy between the immiscible liquids [8]. This interfacial particle film prevents the liquid structures from coarsening as a result of continued phase separation, and generates porous liquid-liquid network structures with surface areas up to 2 $m^2/cm^3$ [20]. Furthermore, STrIPS enables the dispersion of tens of weight percent nanoparticles into the membrane precursor solution yielding a densely particle-covered membrane surface [17], [21]. This facilitates the functionalization of the membrane to tailor the separation selectivity or to impart catalytic properties [17], [21], [22]. However, the fabrication of nanocomposite membranes via STrIPS has so far only been realized for hollow fiber membranes [17], [21], [22].

In this work, we report the continuous fabrication of flat-sheet nanocomposite membranes via STrIPS. We designed a roll-to-roll (R2R) process to produce STrIPS nanocomposite membranes in a steady and controlled fashion. The R2R processing is an established manufacturing technique in which a flexible substrate is passed through an assembly line, followed by coating and modification steps on that substrate [23]. This gives the R2R process beneficial properties such as high-throughput printing, robustness, and cost-effectiveness [24].

In our R2R-membrane fabrication we feed the membrane precursor onto a carrier substrate and subsequently initiate the STrIPS process by pulling the substrate through a water basin. We show that this design allows for the collection of uniform flat-sheet membranes with tunable precursor deposition rate and membrane dimensions. The dynamics of the precursor-substrate interaction are studied by contact angle measurements. Moreover, control over the STrIPS membrane structure is demonstrated by varying the membrane precursor composition. To this end, we analyze the structure of the STrIPS membrane with confocal microscopy and reveal the spatial distribution of nanoparticles and precursor liquids in the membrane.

## 2. Results and discussion

2.1 R2R membrane fabrication via STrIPS

STrIPS nanocomposite membranes are fabricated via liquid-liquid phase separation of a homogeneous precursor mixture. The precursor is composed of the liquids 1,4-butanediol diacrylate (BDA), water and ethanol. Ethanol allows for the mixing of BDA and water above the binodal line of the ternary liquid phase diagram in Fig. 1A [21]. Ludox TMA nanoparticles are dispersed in the liquid mixture using hexadecyltrimethylammonium cations ($CTA^+$) as surfactant (Fig. 1B) [18]. $CTA^+$ can electrostatically adsorb on the surface of the nanoparticles and in this way render the particles interfacially active. For STrIPS membrane fabrication precursor mixtures with liquid compositions *i*, *ii*, and *iii* are prepared as depicted in the ternary phase diagram (Fig. 1A; for detailed compositions see Experimental section).

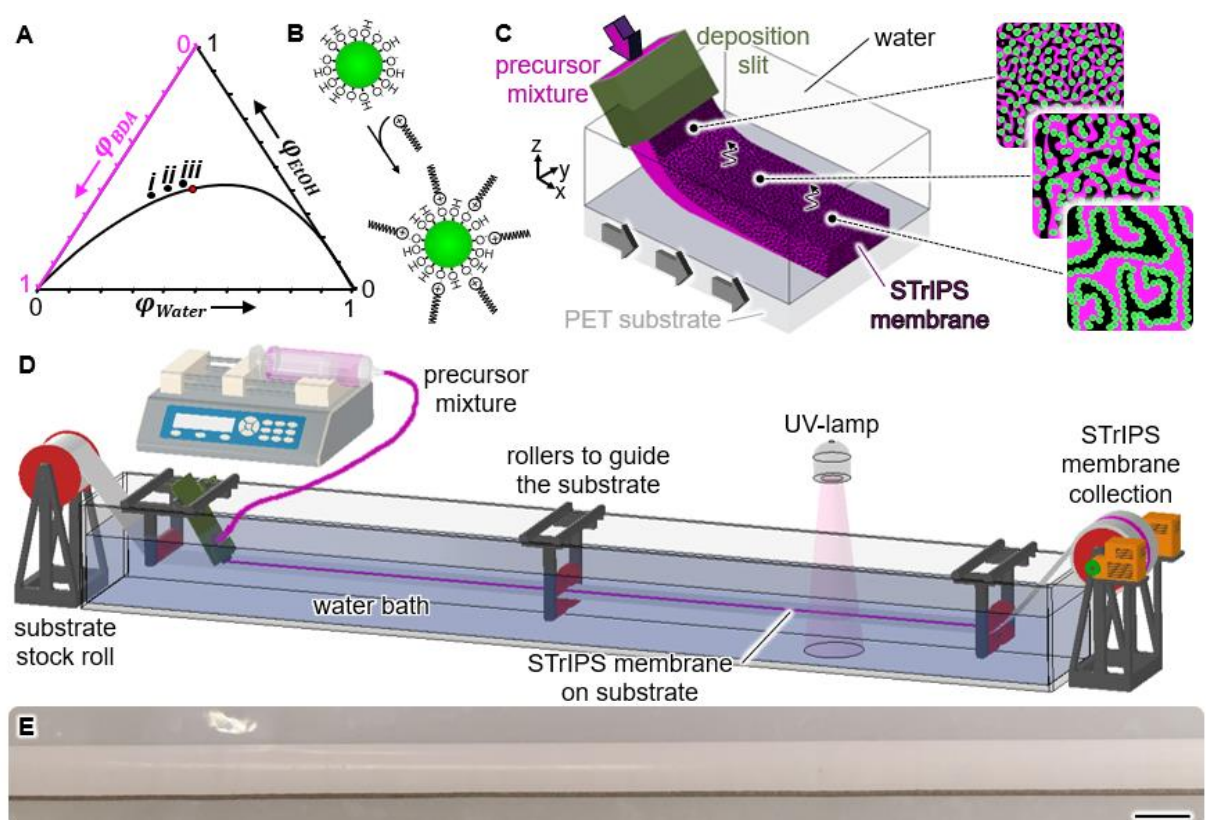

**Figure 1: STrIPS membrane fabrication. A** Ternary phase diagram of 1,4-butanediol diacrylate (BDA), water, and ethanol (EtOH) with liquid volume fractions *φ*. The red dot on the binodal curve gives the critical point. The precursor liquid compositions are shown as *i*, *ii*, and *iii*. **B** Schematic depiction of the electrostatic adsorption of $CTA^+$ on the silica nanoparticles. **C** Cutaway drawing showing the deposition of the precursor mixture via a deposition slit onto polyethylene terephthalate (PET) substrate. The grey arrows indicate the direction of movement of the substrate. Solvent diffusion into the surrounding water initiates the BDA/water demixing which is stabilized by the interfacial

attachment of the nanoparticles. Black color represents water, magenta color BDA and green color silica nanoparticles. **D** Schematic representation of the roll-to-roll process for the continuous fabrication of flat-sheet nanocomposite membranes via STrIPS. **E** Photograph of the STrIPS membrane (white) on PET substrate (transparent). Scale bar is 10 mm.

To initiate STrIPS, the precursor mixture is pumped through a rectangular slit of 100 µm height into a water reservoir (pH 3; Fig. 1C; for details see SI section S2). As the precursor flows out, ethanol diffuses into the surrounding water and triggers the phase separation of BDA and water. The $CTA^+$-modified nanoparticles stabilize the phase separation by adsorbing at and rigidifying the BDA/water interface (Fig. 1C) [21]. The interfacial assembly of the particles is facilitated after substantial amounts of ethanol have diffused out of the liquid mixture as the particle adsorption energy is lowered by the rise in interfacial tension between BDA and water [8], [25], [26].

For the continuous production of nanocomposite membranes, we combine the process of the precursor mixture undergoing STrIPS with the R2R-film fabrication technique. The precursor mixture is therefore deposited onto a flexible carrier substrate which is continuously pulled through the R2R-setup (Fig. 1D). As substrate, polyethylene terephthalate (PET) is unrolled from a stock roll into a water bath. In the water bath, rollers guide the substrate closely underneath the precursor deposition slit where the precursor mixture flows out onto the substrate. Upon deposition, the diffusion of ethanol out of the precursor mixture initiates STrIPS. The BDA/water phase separation proceeds while the precursor travels through the water bath. The resulting particle-stabilized emulsion gel is solidified via UV-light polymerization of the BDA. To this end, 3-5 wt % 2-hydroxy-2-methylpropriophenone (HMPP) is added to the precursor. HMPP generates radicals to form cross-linked polyBDA once the STrIPS membrane passes under the UV-lamp located at the end of the water bath. Finally, the polymerized STrIPS membrane is reeled with the PET substrate on a collection roll.

The precursor mixture needs to wet the PET substrate to fabricate uniform STrIPS membranes. The substrate wetting can be expressed via the three-phase contact angle $\theta$ between precursor mixture, PET substrate and water (pH 3) as the continuous phase (Fig. 2A). Following this, we measured the contact angle of a particle-free precursor droplet with liquid composition *i*, *ii*, and *iii* containing 50 mM $CTA^+$ dispensed on PET in water (for exact droplet composition see Experimental section). To diminish the effect of gravitation on the droplet deformation, the droplet volume is kept at 5 µL for all precursors. The photographs in Fig. 2A show that immediately after deposition precursor *i* balls up on the PET substrate to $\theta_{initial} = 109 (\pm 8)°$, and precursor *ii* to $\theta_{initial} = 103 (\pm 3)°$ while precursor *iii* spreads on the PET to $\theta_{initial} = 68 (\pm 8)°$. Thus, the initial wetting of the PET in water improves from precursor *i* to *iii*.

The wetting depends on the liquid composition of the precursor droplet and the interplay with the continuous phase and the substrate [27], [28]. Regarding the liquid starting composition of the droplets, the BDA volume fraction decreases from precursor *i* ($\varphi_{BDA} = 0.433$) to *ii* ($\varphi_{BDA} = 0.367$) to *iii* ($\varphi_{BDA} = $ 0.303). Precursor mixtures with lower BDA volume fractions naturally have higher water/EtOH contents such that they can have stronger hydrophilic interactions with the surrounding water phase. This potentially favors the instant enlargement of the surface area between precursor and surrounding water. Furthermore, EtOH as solvent and $CTA^+$ as surfactant can contribute to the wetting by lowering the interfacial energy at the precursor/water and the precursor/PET contact lines.

Simultaneously to the dispensing of the precursor droplet, ethanol starts to diffuse out into the continuous phase and causes the precursor liquids inside the droplet to phase separate. For this stage, we observe nucleation of BDA-rich microdroplets as a dark blur around the precursor droplet/water contact line (Fig. 2B). With ongoing phase separation the water dissolves from the precursor into the continuous phase. Although for STrIPS membrane fabrication the nanoparticles arrest the phase separation before complete demixing of BDA and water, from previous research it is known that liquid-liquid phase separation affects the wetting of a substrate e.g. by spontaneous nucleation events, changing surface interactions upon coarsening of the liquid/liquid domains or hydrodynamic flows [29]–[31].

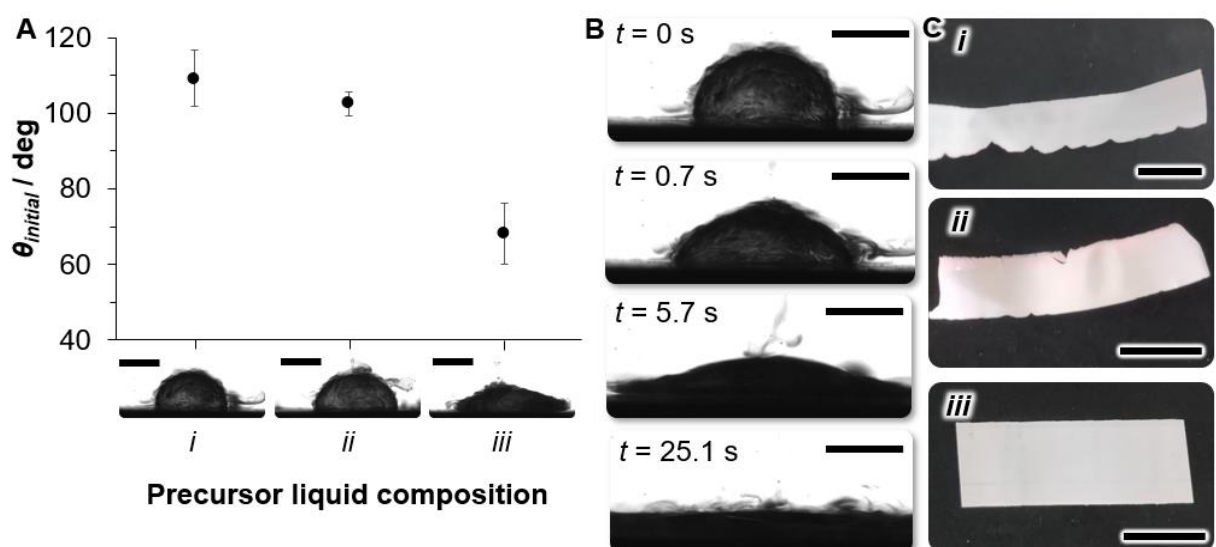

**Figure 2: Substrate wetting by the precursor mixture. A** Initial three-phase contact angle $\theta_{initial}$ of a precursor droplet of liquid composition *i*, *ii*, and *iii* dispensed on polyethylene terephthalate (PET) substrate in water (pH 3). The insets show photographs of the precursor droplets directly after deposition on PET. **B** Snapshots of precursor *i* spreading on PET in water (pH 3) over time *t*. Scale bars are 2 mm. **C** Photographs showing the STrIPS membrane shape after fabrication from precursor mixtures of liquid composition *i*, *ii*, and *iii*. Scale bars are 10 mm.

As shown in Fig. 2B, the precursor droplet will spread over the PET within a few seconds. Potentially, compositional changes induced by ethanol diffusion and BDA/water demixing change the surface interactions between precursor, surrounding water phase, and PET and promote the wetting of the substrate over time. After completion of the phase separation the BDA spreads and covers the PET. $CTA^+$ can favor the spreading of the BDA by lowering the surface interactions towards PET and water.

However, also pure BDA spreads on PET in the absence of any surfactant (see SI section S4). Moreover, the diffusion of ethanol from the droplet into the surrounding water can cause lateral convective flows that induce spreading of the phase separating precursor droplet [29]. However, more detailed research is required to investigate the mechanisms of the wetting of the PET substrate by this ternary liquid mixture undergoing phase separation.

Furthermore, in the fabrication of uniform STrIPS membranes the thickness of the membrane can be controlled. The membrane thickness depends on the speed at which the PET substrate is moving while the precursor is deposited, and the flow rate at which the precursor is coated onto the substrate. For increasing the pulling speed of the substrate from 55 mm/s to 148 mm/s the thickness of the STrIPS membrane decreases from 90 (± 10) µm to 31 (± 9) µm at a precursor flow rate of 3 mL/min (Fig. 3A). When the precursor flow rate is increased from 2.2 mL/min to 3.8 mL/min the STrIPS membrane thickness changes from 45 (± 6) µm to 90 (± 25) µm for a substrate pulling speed of 70 mm/s (Fig. 3B). As STrIPS can proceed through the entire depth of the membrane the resulting structure is unaffected by the change in thickness.

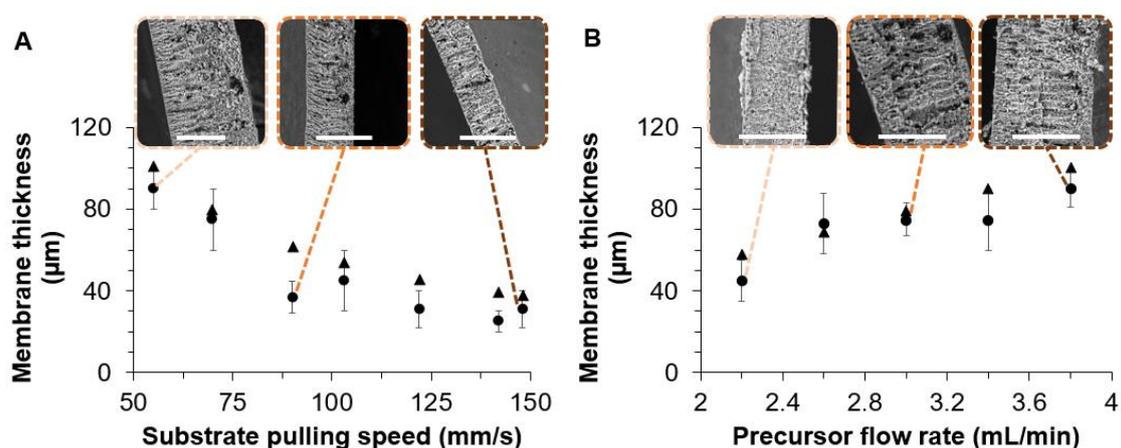

**Figure 3: STrIPS membrane thickness control. A** Thickness of the STrIPS membrane at different pulling speed of the PET substrate with scanning electron microscopy (SEM) micrographs of the membranes above. The round points (●) give the measured thickness, the triangular points (▲) the calculated thickness. The error bars give the standard deviation from measuring the thickness of five different membrane cross-sections. The precursor flow rate is kept at 3 mL/min. **B** Thickness of the STrIPS membrane at different precursor flow rates and a constant substrate pulling speed of 70 mm/s. The round points (●) give the measured thickness, the triangular points (▲) the calculated thickness, with error bars for the standard deviation of the thickness measured from five membrane cross-sections. Above SEM micrographs of the resulting STrIPS membrane are shown. All scale bars are 50 µm.

The thickness of the STrIPS membrane can also be calculated based on the assumption that the volume of precursor that flows through the deposition slit is recovered on the PET substrate (for details see SI section S5). Overall, the calculated membrane thickness is higher, but follows the trend observed in the experimental determination of the thickness (Fig. 3A+B). This discrepancy probably arises from changes in the mass of the membrane during STrIPS due to the outward diffusion of ethanol and the potential inflow of water [20]. Likewise, the spreading of the precursor mixture on PET broadens the membrane but is disregarded in the calculation. The substrate pulling speed and the precursor flow rate nevertheless regulate the amount of precursor that is deposited on the substrate and thus control the thickness of the STrIPS membrane.

2.2 STrIPS membrane structure control

The STrIPS membrane structure depends on the liquid starting composition of the precursor mixture. To analyze the membrane structure for the spatial distribution of BDA, water, and nanoparticles, confocal laser scanning microscopy is employed. To this end, fragments of the STrIPS membrane are cut out from the collection roll, dried at room temperature, and removed from the PET substrate by detachment during drying. The membrane samples are prepared for confocal microscopy as described in [21]. In brief, the dye Nile red is added to the precursor mixture to fluorescently label the BDA-rich phase. After UV-polymerization, the STrIPS membrane is submerged in an alkaline solution of the fluorescent dye Rhodamine 110 chloride, allowing for the adsorption of Rhodamine 110 onto the nanoparticles. For spatially resolved structure analysis, the membrane is immersed in diethyl phthalate, a liquid with a refractive index close to polyBDA ($n \approx 1.50$) [32]. The Nile red is excited with 561 nm laser light with the fluorescence detected in the wavelength range of 600-700 nm; Rhodamine 110 is excited at 488 nm laser light and the fluorescence is detected at 500-550 nm (see SI section S6).

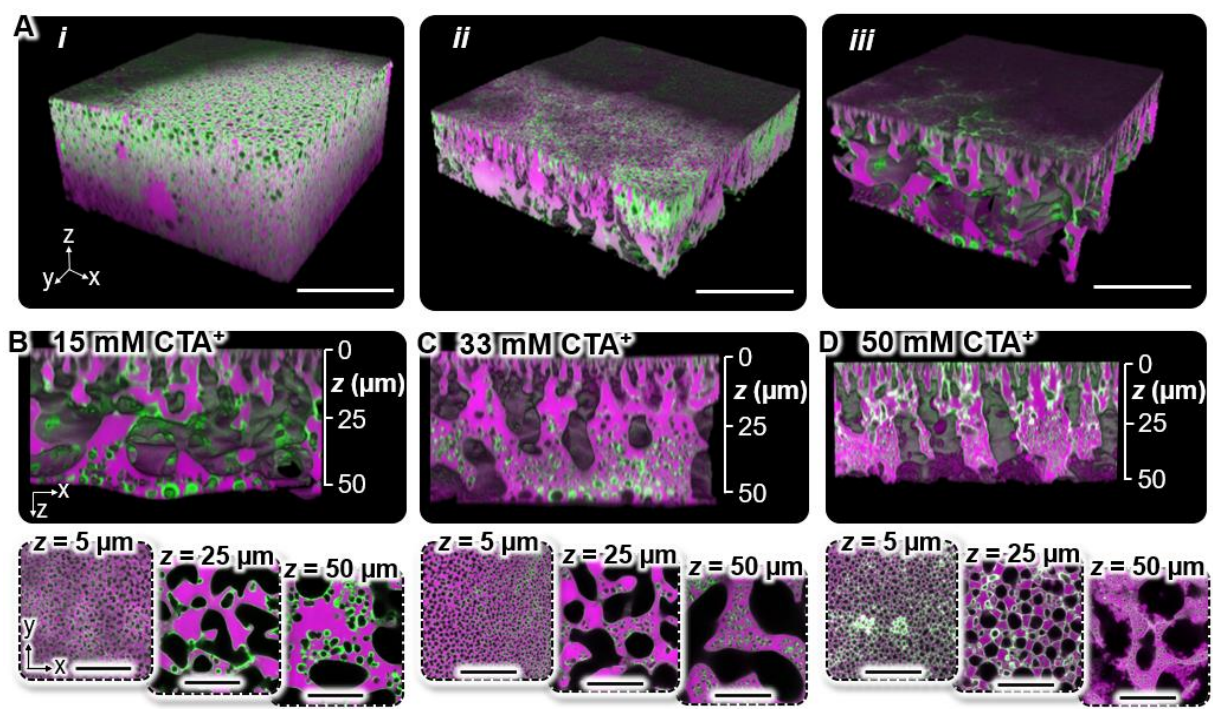

**Figure 4: STrIPS membrane structure control. A** Three-dimensional reconstructions of confocal micrographs of STrIPS membranes made with precursor compositions *i*, *ii*, and *iii*. All precursor mixtures contain 15 mM CTA$^+$ and 11 wt % nanoparticles. Black color represents water, magenta color polyBDA and green color silica nanoparticles. White scale bars are 50 µm. **B** Cross-section of the STrIPS membrane fabricated from precursor *iii* with 11 wt % nanoparticles and 15 mM CTA$^+$. Below,

confocal slices from different membrane depths $z$ are shown. **C** Same as B using 33 mM $CTA^+$. **D** Same as B using 50 mM $CTA^+$. Black scale bars are 25 µm.

According to the three-dimensional confocal reconstructions in Fig. 4A the structure of the STrIPS membrane changes with the liquid starting composition of precursor *i*, *ii*, and *iii*. The STrIPS membrane fabricated from precursor *i* is composed of water compartments embedded in polyBDA. This structure has likely formed via nucleation and growth of water domains in BDA during STrIPS. The water domains are enclosed by a film of nanoparticles indicating that the particles stabilized the BDA/water demixing. Compared to precursor *i*, precursors *ii* and *iii* result in STrIPS membranes with higher porosity, with water domains coarsening from the top to the bottom of the membrane. The proximity of the precursor liquid compositions *ii* and *iii* to the critical point of the ternary phase diagram (Fig. 1A) suggests that for these precursors the BDA/water phase separation has proceeded via spinodal decomposition. In our previous work we have shown that for this BDA/water/ethanol precursor system the nanoparticles fail to stabilize the spinodal decomposition instantaneously, resulting in a coarsening of the water domains before an interfacial film of nanoparticles stops the liquids from further demixing and arrests the membrane structure [21].

The coarsening of the water domains from the surface to the bottom of the membrane can be explained via the solvent diffusion dynamics during STrIPS. We established a model exclusively considering the diffusion of ethanol out of the membrane and tracked the structural evolution of the membrane during STrIPS. A detailed explanation of the model and all findings is provided in the Supporting Information (section S7). The model shows that the ethanol concentration decreases at the top surface of the membrane immediately when the precursor is deposited on the substrate (Fig. S8). In consequence, the nanoparticles can stabilize the BDA/water phase separation at the membrane surface at an early stage of STrIPS, resulting in small membrane surface pores. Below the surface the ethanol concentration remains elevated over a longer time, delaying the stabilization of the BDA/water phase separation inside the membrane. According to the experimental investigation of the structural evolution, for a 50 µm thick membrane it takes around five seconds until the BDA/water phase separation has been arrested also at the bottom of the membrane (Fig. S9). The gradual proceeding and stabilization of STrIPS from the surface to the bottom of the membrane leads to a coarsening of the membrane pores.

The $CTA^+$ concentration in the precursor liquid has a pronounced effect on the spinodal arrangement of BDA and water. Upon increasing the $CTA^+$ concentration from 15 mM to 50 mM in the precursor mixture, the polyBDA domains of the membrane seem to curve stronger around the water pores (Fig. 4B-D). Deeper inside the membrane, higher $CTA^+$ concentrations result in the formation of small water droplets embedded in the polyBDA domains ($z = 50$ µm; Fig. 4B-D). This curving of the polyBDA domains around the water pores suggests that for elevated $CTA^+$ concentrations the particles have partitioned into the BDA-rich phase before arresting the BDA/water phase separation.

At higher CTA$^+$ concentrations more surfactant can adsorb and hydrophobize the particle surface [33]. As hydrophobic particles preferentially migrate into the BDA-rich phase, the particles can stabilize the growth of water droplets in the BDA-rich domains [21]. In addition, less particle aggregates can be found in the water pores of the membrane fabricated with 33 mM and 50 mM CTA$^+$ compared to the membrane with 15 mM CTA$^+$ (Fig. 3B). This supports the hypothesis that the particles preferentially partition into the BDA-rich phase for higher CTA$^+$ concentrations so that less particles are available to aggregate in the water pores. As a result of the particle hydrophobization, the particles impose an interfacial curvature around the water pores and favor the formation of nucleated water sub-structures inside the BDA-rich domains.

However, at the bottom of the membrane fabricated with 50 mM CTA$^+$ also polyBDA droplets inside the water domains can be found ($z$ = 50 µm; Fig. 3D). These emulsified polyBDA-in-water droplets have likely been stabilized by more hydrophilic particles [21], [34]. The prolonged high ethanol concentration at the bottom of the membrane during STrIPS can reduce the affinity of the CTA$^+$ to adsorb on the particles due to better solubility of the surfactant [35]. This renders the particles less hydrophobic and favors the dispersibility of the particles in the water-rich phase and so the emulsification of polyBDA droplets at the bottom of the membrane. Hence, the membrane pore morphology is determined by the interplay of CTA$^+$ and ethanol concentration during STrIPS.

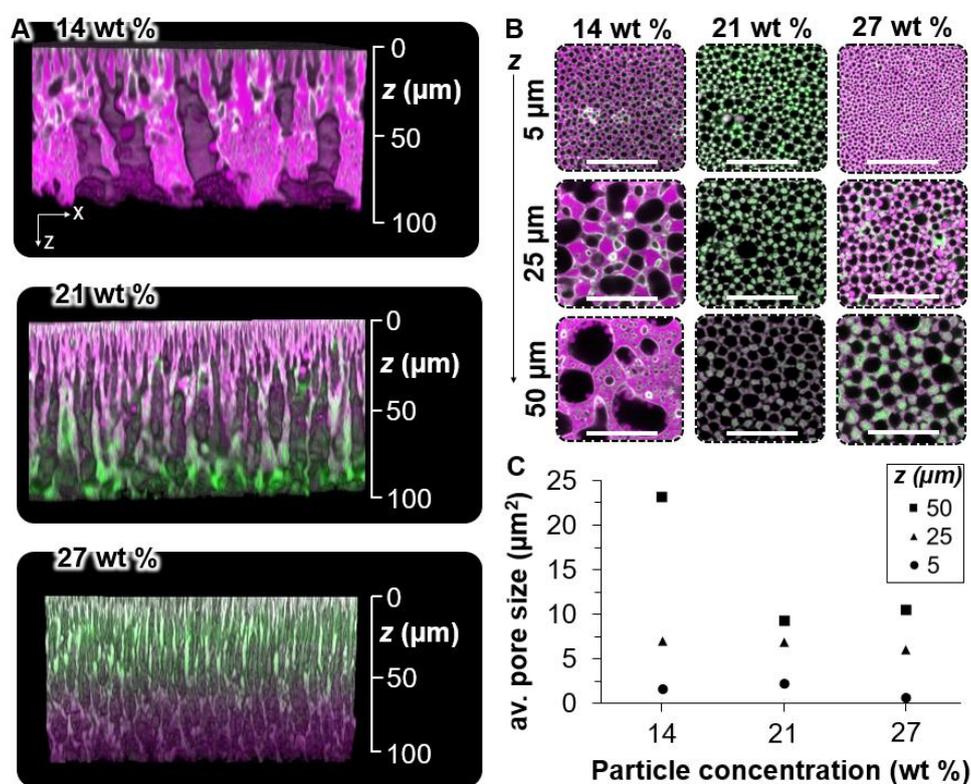

**Figure 5: STrIPS membrane porosity. A** Cross-sections of STrIPS membranes fabricated from precursor liquid composition *iii* using 50 mM CTA$^+$ and different nanoparticle concentrations. **B** Confocal micrographs of membrane slices from different depths *z*. All scale bars are 50 µm. **C** Average (av.) pore sizes at membrane depths *z* for the STrIPS membranes fabricated with different nanoparticle concentrations. The average pore sizes correspond to the confocal micrographs shown in B.

The morphology of the membrane pores overall remains irregular due to the coarsening of the BDA/water domains over the depth of the membrane. However, we find that the membrane pores can be deformed into tubular structures which stretch from the surface to the bottom of the membrane by increasing the particle concentration in the precursor mixture (Fig. 5A). While the membrane fabricated with 14 wt % particles in the precursor mixture has broad and irregularly shaped pores, the water pores are narrowed towards a tubular shape for particle concentrations of 21 wt % and 27 wt % in the precursor (Fig. 5A). Thus, high particle concentrations in the precursor mixture allow for the preparation of STrIPS membranes with pores extending across the membrane depth.

Increasing the particle concentration in the precursor mixture also reduces the coarsening of the water pores deeper inside the membrane. The confocal micrographs in Fig. 5B show that higher particle concentrations result in smaller membrane pores especially underneath the membrane surface. The surface pores (at $z = 5$ µm to exclude uneven surface regions) shrink from an average size of 1.7 µm$^2$ at 14 wt % particles and 2.2 µm$^2$ at 21 wt % particles to 0.7 µm$^2$ at 27 wt % particles (Fig. 5C). In the center of the membrane ($z = 50$ µm) the pores are narrowed from 23.2 µm$^2$ (14 wt % particles) to 9.3 µm$^2$ (21 wt % particles) and 10.5 µm$^2$ (27 wt % particles). The narrowing of the pores implies that the coarsening of the BDA/water domains is stabilized faster with increasing particle concentration. For higher particle concentrations in the precursor mixture more particles are available to arrest the BDA/water phase separation. Before the phase separation is arrested, the water pores potentially fuse, and form elongated tubular structures across the depth of the STrIPS membrane.

## 3. Conclusions

In this work, we introduce a roll-to-roll (R2R)-approach for the continuous fabrication of porous flat-sheet nanocomposite membranes via solvent transfer induced phase separation (STrIPS). To this end, we print a precursor mixture composed of butanediol diacrylate (BDA), water, and ethanol with CTA$^+$-functionalized Ludox TMA nanoparticles onto polyethylene terephthalate (PET). The precursor is continuously deposited on the PET which moves as carrier substrate through a water bath to trigger the membrane formation via STrIPS. To produce membranes of controlled dimensions, the wetting dynamics of the precursor mixture on PET are studied by contact angle measurements. Confocal microscopy analysis reveals the spatial distribution of BDA, water, and nanoparticles inside the membrane. We find that the precursor liquid composition and the CTA$^+$-modification of the particles change the arrangement of the BDA/water domains and the membrane porosity. Furthermore, increasing the nanoparticle concentration in the precursor deforms the membrane pores to narrow tubular structures extending through the depth of the membrane. Our R2R-process enables the fabrication of flat-sheet nanocomposite membranes with defined structure, dimensions, and pore morphology via STrIPS. This opens the potential for further advancements in the field of STrIPS membrane synthesis.

## 4. Experimental section

*4.1 Membrane precursor preparation*

The membrane precursor consists of Ludox TMA nanoparticles (spherical, particle diamater 20 nm; Grace) dispersed in a homogeneous mixture of 1,4-butanediol diacrylate (BDA), water, and ethanol (EtOH). The nanoparticle dispersion is prepared by concentrating 50 mL Ludox TMA from 34 wt % to 40 wt % by the evaporation of water (Rotary evaporator, Heidolph Instruments) at 60 °C and 140 mbar. The concentrate is centrifuged at 3750 rpm for 10 min (Allegra X-12R, Beckman Coulter) to remove particle aggregates and adjusted to pH 3 by the addition of 1 M HCl (Acros Organics).

We start the preparation of the precursor mixtures *i*, *ii*, and *iii* by mixing the pure liquids BDA, water (MilliQ purification system), and EtOH: *i* ($\varphi_{BDA} = 0.433$; $\varphi_{Water} = 0.032$; $\varphi_{EtOH} = 0.369$); *ii* ($\varphi_{BDA} = 0.367$; $\varphi_{Water} = 0.040$; $\varphi_{EtOH} = 0.387$); *iii* ($\varphi_{BDA} = 0.303$; $\varphi_{Water} = 0.051$; $\varphi_{EtOH} = 0.395$). Hexadecyltrimethylammonium cations ($CTA^+$; Sigma Aldrich) are dissolved as 200 mM stock solution in ethanol and added to the mixtures *i* ($\varphi_{200\ mM\ CTA+} = 0.030$), *ii* ($\varphi_{200\ mM\ CTA+} = 0.031$), and *iii* ($\varphi_{200\ mM\ CTA+} = 0.032$) to obtain 15 mM $CTA^+$ in the precursor mixture. The concentrated Ludox TMA dispersion is added to precursor *i* ($\varphi_{Ludox\ TMA} = 0.136$), *ii* ($\varphi_{Ludox\ TMA} = 0.175$), and *iii* ($\varphi_{Ludox\ TMA} = 0.219$) yielding a particle concentration of 11 wt % in the precursor mixture. The photo-initiator 2-hydroxy-2-methylpropiophenone (Sigma Aldrich) is incorporated to 3-5 wt % to the precursor mixture to polymerize the BDA monomers after phase separation (for details see SI section S1). Visualization of the BDA-rich phase is accomplished by fluorescent labelling with Nile red (Sigma Aldrich). All chemicals are purchased as analytical grade.

*4.2 Roll-to-roll membrane fabrication*

The precursor mixture is pumped at a rate of 2-4 mL/min through a rectangular slit of 8 mm width and 100 μm height onto polyethylene terephthalate foil (PET; 100 μm thick; Reflectiv). The precursor deposition slit is built from three microscopy cover slips (20 x 20 x 0.1 mm; Menzel-Gläser) as shown in detail in SI section S2. All inner glass surfaces of the slit are coated with a solution of 0.2 wt % poly(diallyldimethylammonium chloride) (PDADMAC; Sigma Aldrich) and 500 mM NaCl (Merck) to prevent adhesion of the precursor mixture. The precursor deposition slit is positioned at an angle of 55° with respect to the PET substrate. The substrate is pulled with two electric motors (LEGO Technic XL Motor) at a default speed of 5 cm/s through the R2R-setup. The R2R-setup is submerged in a container of 100 cm length filled with 9 L demineralized water (MilliQ purification system) which is brought to pH 3 by addition of 1 M HCl. For UV-polymerization the membrane is exposed to high intensity UV-light using a UV-lamp with focused beam (OmniCure Series 1500).

The rollers guiding the substrate through the R2R-machine are 3D-printed with polylactic acid filament (Monoprice 3D printer filament; Dremel DigiLab). The holders for the rollers, the substrate stock roll and the collection roll are built with Lego bricks (for details see SI section S3).

*4.3 Membrane structure characterization*

The STrIPS membrane structure is analyzed by confocal laser scanning microscopy (Stellaris 5, Leica Microsystems). After polymerization, the STrIPS membrane is washed in a solution of 50 vol-% 1 M HCl and 50 vol-% ethanol to remove $CTA^+$. The membrane is then immersed in an alkaline Rhodamine 110 (Chemodex) solution and made optically transparent by replacing water with diethyl phthalate (Acros Organics). Upon excitement with 488 nm laser light the Rhodamine 110 labelled particles emit green fluorescence detected at 500-550 nm. The Nile red fluorescence of the polyBDA is excited with 561 nm laser light and detected at 600-700 nm. The membrane surface and cross-section is imaged via scanning electron microscopy (Phenom ProX, Thermo Fisher Scientific) applying an electron beam excitation of 10 kV and an 8 nm layer of sputter-coated platinum.

For the determination of the STrIPS membrane thickness, a precursor mixture of the liquids BDA, water, and methanol (MeOH) with liquid composition $\varphi_{BDA} = 0.40$, $\varphi_{Water} = 0.16$, and $\varphi_{MeOH} = 0.44$ is used containing 11 wt % nanoparticles and 15 mM $CTA^+$. The precursor flow rate is adjusted via a syringe pump (World Precision Instruments) and the substrate velocity is controlled by the motor pulling speed (LEGO Technic XL Motor). The cross-section of the STrIPS membrane is imaged via SEM and the membrane thickness is measured from five membrane pieces of a defect-free structure.

*4.4 Contact angle measurements*

Particle-free precursor mixtures of liquid composition *i*, *ii*, and *iii* are prepared containing 50 mM $CTA^+$. For all mixtures the nanoparticle volume fraction $\varphi_{Ludox\ TMA}$ is replaced by water. 5 µL of each precursor liquid is deposited on PET substrate in a glass cylinder filled with water of pH 3 (Fig. S5). The droplet behavior is observed and recorded from the side-view with a camera (25 fps; Thorlabs CS165MU/M). All images are analyzed with the software Fiji ImageJ (version 1.53k14). The dynamical apparent contact angle is averaged from three different side-view images by ellipsoid fitting.

*4.5 Pore size analysis*

The membrane pore sizes are measured from analyzing confocal micrographs with the software Fiji Image J (version 1.53k14). As explained in detail in SI section S8, the image contrast is enhanced and pixel thresholding is performed to accurately highlight the water pores. After binarization of the image the pore area is measured via the "Analyze particle" feature. To confirm that the membrane pores have correctly been identified by the software, the pore outlines are placed over the original confocal micrograph.

**Notes**

The authors declare no competing financial interest.

**Author information**

**Corresponding author**

**M. F. Haase** - Van't Hoff Laboratory of Physical and Colloid Chemistry, Department of Chemistry, Debye Institute for Nanomaterials Science, Utrecht University, Utrecht, The Netherlands; Email: m.f.haase@uu.nl

**Authors**

**H. Siegel** - Van't Hoff Laboratory of Physical and Colloid Chemistry, Department of Chemistry, Debye Institute for Nanomaterials Science, Utrecht University, Utrecht, The Netherlands

**M. de Ruiter** - Van't Hoff Laboratory of Physical and Colloid Chemistry, Department of Chemistry, Debye Institute for Nanomaterials Science, Utrecht University, Utrecht, The Netherlands

**C. M. Hesseling** - Van't Hoff Laboratory of Physical and Colloid Chemistry, Department of Chemistry, Debye Institute for Nanomaterials Science, Utrecht University, Utrecht, The Netherlands

**G. Athanasiou** - Van't Hoff Laboratory of Physical and Colloid Chemistry, Department of Chemistry, Debye Institute for Nanomaterials Science, Utrecht University, Utrecht, The Netherlands

**Acknowledgement**

This publication is part of the project "Bijel templated membranes for molecular separations" (with project number 18632 of the research program Vidi 2019) which is financed by the Dutch Research Council (NWO). We thank Jesse M. Steenhoff for kind support in the transient solvent diffusion modelling.

# Supporting Information

# Scalable fabrication of bijel films via continuous flow slit-coating

*Henrik Siegel[1,‡], Mariska de Ruiter[1,‡], Cos M. Hesseling[1], Georgios Athanasiou[1], Martin F. Haase[1]\**

[1] Van't Hoff Laboratory of Physical and Colloid Chemistry, Department of Chemistry, Debye Institute for Nanomaterials Science, Utrecht University, Utrecht, The Netherlands

‡ Both authors contributed equally

\* Correspondence: m.f.haase@uu.nl

**Table of contents**

## S1. Membrane precursor preparation

Homogeneous STrIPS membrane precursor dispersions are prepared by mixing 1,4-butanediol diacrylate (BDA) with water and ethanol (EtOH) at compositions *i*, *ii*, and *iii* given in Fig. S1. Miscibility of the precursor mixtures is dictated by the binodal curve of the ternary phase diagram taken from [1]. The critical point for this ternary liquid system is $\varphi_{BDA} = 0.301$, $\varphi_{EtOH} = 0.427$, and $\varphi_{Water} = 0.272$ [1]. To all liquid mixtures, Ludox TMA nanoparticles (40 wt % aqueous dispersion, pH 3) are added to yield a total particle concentration of 11 wt % in the precursor mixture. Hexadecyltrimethylammonium cations ($CTA^+$) are dissolved in EtOH and added as 200 mM stock solution to the precursor mixture to obtain a total of 15 mM $CTA^+$ in the membrane precursor. Small amounts of the photo-initiator 2-hydroxy-2-methylpropriophenone (HMPP) are dissolved in the precursor for the polymerization of the BDA.

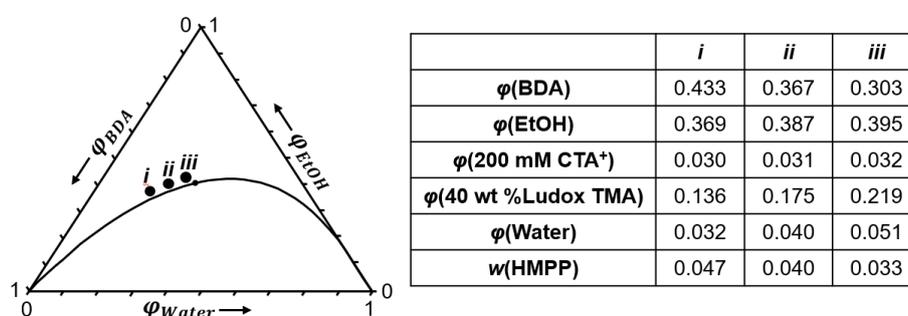

|  | *i* | *ii* | *iii* |
|---|---|---|---|
| $\varphi$(BDA) | 0.433 | 0.367 | 0.303 |
| $\varphi$(EtOH) | 0.369 | 0.387 | 0.395 |
| $\varphi$(200 mM $CTA^+$) | 0.030 | 0.031 | 0.032 |
| $\varphi$(40 wt %Ludox TMA) | 0.136 | 0.175 | 0.219 |
| $\varphi$(Water) | 0.032 | 0.040 | 0.051 |
| $w$(HMPP) | 0.047 | 0.040 | 0.033 |

**Figure S1:** Ternary phase diagram for BDA/water/EtOH with STrIPS membrane precursor compositions *i*, *ii*, and *iii*. Liquid volume fractions are expressed by $\varphi$, and $w$ represents the HMPP weight fraction in the precursor mixture.

## S2. Precursor deposition device

The precursor deposition device is built from three microscopy cover slips (20 x 20 x 0.1 mm; Menzel-Gläser; Fig. S2-A). Two microscopy cover slips are glued together with UV-glue (Norland Optical Adhesive 81) including one cover slip which is cut into strips of 6 mm width (Sutter scoring tile) as spacer between them. The precursor mixture is supplied via PTFE tubing (Cole Parmer 20" Thin Wall Tubing Natural). To connect the tubing with the slide assembly, a ~20 mm long slit is cut into the PTFE tubing with a razor blade. The slide assembly is pushed into the slit and both parts are sealed with Epoxy glue (Liqui Moly 5 Minute Epoxy Glue) as shown in Fig. S2-B.

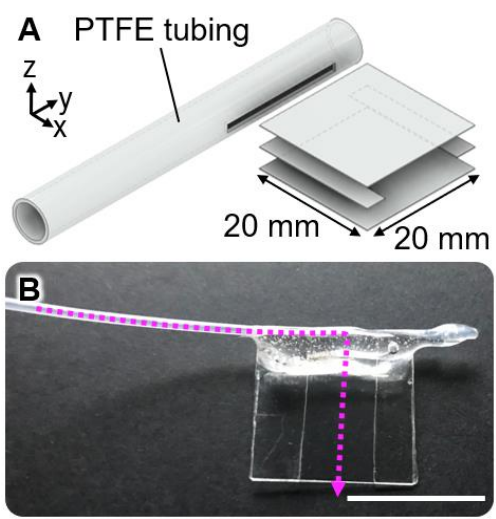

**Figure S2: A** Schematic of the precursor deposition device composed of microscopy glass slides and tubing. **B** Photo of the precursor deposition device. The precursor mixture is pumped through PTFE tubing and exits through the slit between the microscopy glass slides. The deposition path of the precursor mixture is marked in magenta. White scale bar is 20 mm.

To be uniformly deposited onto the substrate, the precursor mixture needs to have a unidirectional flow profile at the opening of the deposition slit. The precursor flow dynamics are studied by microfluidic simulations using COMSOL® Multiphysics (version 5.5). One quadrant of the precursor deposition slit is modelled at real-scale dimensions of 22 mm length, 4 mm width and 50 µm height (Fig. S3-A). For the simulations a measured density of the precursor mixture of 1.588 g/mL is used and a dynamic viscosity of 0.00242 Pa*s as determined by rheology measurements. Given a precursor flow rate of 55 mm/s which is used for STrIPS membrane fabrication a Reynolds number of 144 indicates a laminar flow regime. For the numerical calculation of the precursor flow profile a fine mesh is built along the wall of the slit which coarsens towards the center (Fig. S3-A). The simulation is performed selecting a no-slip boundary condition for the walls and a quadratic order discretization for the fluid velocity.

In Fig. S3-B the velocity profile of the precursor mixture is plotted over the width of the deposition slit. Here, $y$ refers to different positions along the length of the slit starting at $y = 0$ mm at the connection to the PTFE tubing and extending to $y = 22$ mm where the precursor flows out of the slit. When the precursor flows through the deposition device a flat flow profile develops along the width of the slit. Likewise, a laminar flow profile evolves across the height of the slit (Fig. S3-C). In both $x$- and $z$-direction the precursor flow profile is fully developed at length $y = 1$ mm of the slit as indicated by the overlap of the fluid velocity profiles for $y \geq 1$ mm. The design of the precursor deposition device thus allows for the uniform deposition of the precursor mixture onto the substrate.

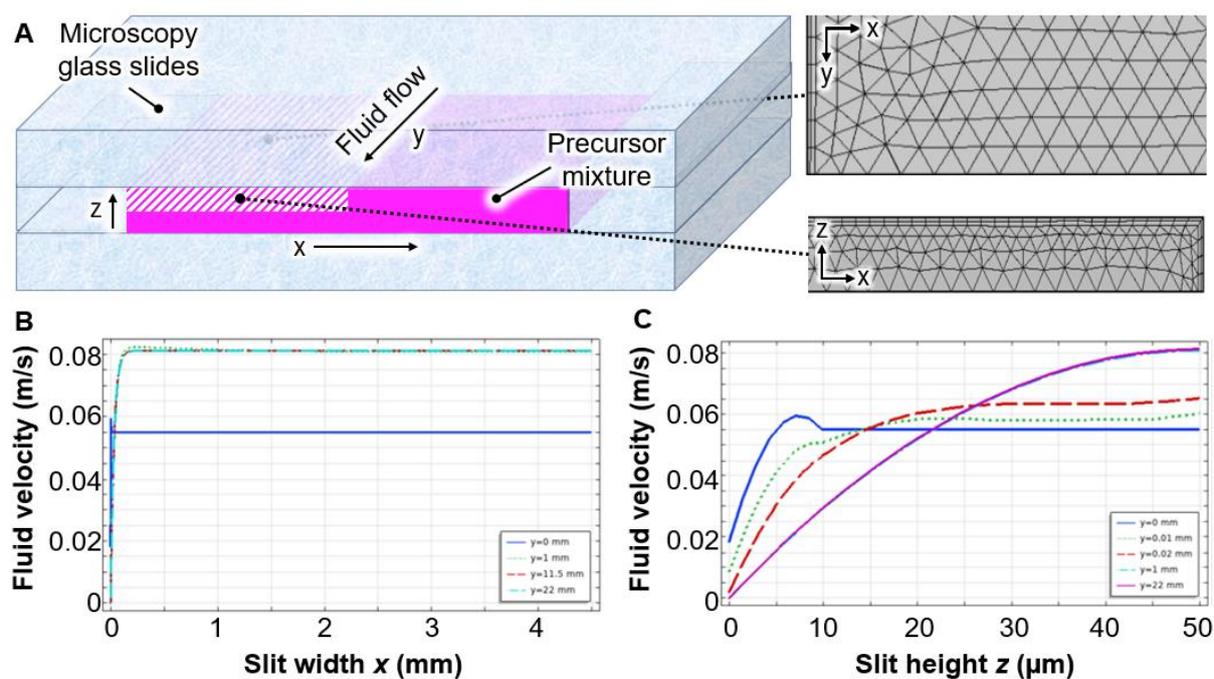

**Figure S3: A** Three-dimensional schematic of the precursor deposition device. The precursor mixture is colored in magenta with the magenta/white dashed region representing the quadrant for which the precursor velocity profile is calculated. The mesh shows the nodes for the numerical calculation of the fluid velocity across the precursor deposition slit in *xy*- and *xz*-direction. **B** Velocity of the precursor mixture over the width *x* of the deposition slit for a precursor flow rate of 55 mm/s. **C** Velocity of the precursor mixture across the height *z* of the deposition slit at a precursor flow rate of 55 mm/s.

## S3. Roll-to-roll machine assembly

STrIPS membranes are fabricated with the roll-to-roll (R2R-) machine in Fig. S4-A: The polyethylene terephthalate (PET) substrate is unrolled from the stock roll (a) and guided closely below the holder (b) for the precursor deposition device. The precursor mixture is supplied via a syringe pump (c) and flows through the deposition device onto the substrate. The substrate coated with precursor is pulled through a container filled with water (d) to initiate STrIPS. In the R2R-machine rollers (e) keep the substrate in position. After passing under a UV-lamp for photo-polymerization (f) the STrIPS membrane is collected on a roll that is driven by two electric motors (g). As shown in Fig. S4-B, the deposition device is placed at an angle of 55⁰ with respect to the substrate. This device configuration can be adjusted via several gears. The R2R-machine allows for the fabrication of uniform STrIPS membranes (Fig. S4-C).

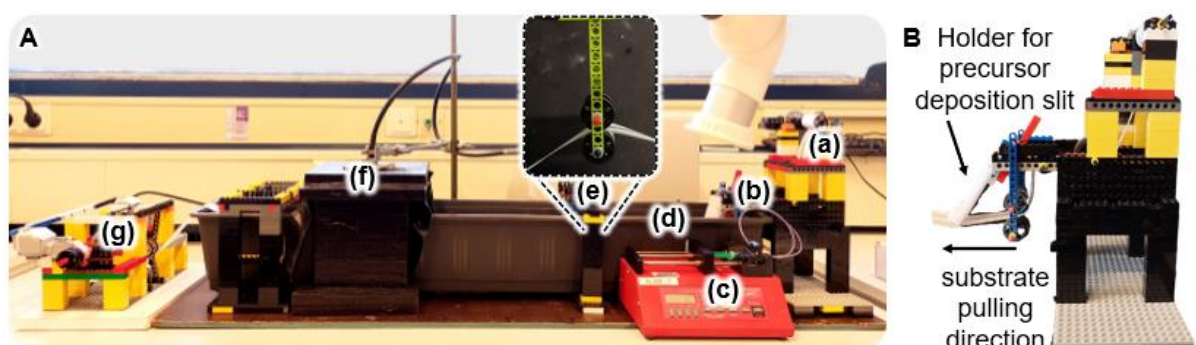

**Figure S4: A** Roll-to-roll machine for the fabrication of STrIPS nanocomposite membranes with the following components: (a) substrate stock roll; (b) holder for the precursor deposition device; (c) syringe pump to flow the precursor mixture; (d) water-filled container; (e) roller to keep substrate in position (detailed view in inset); (f) UV-lamp with housing for the photo-polymerization of the STrIPS membrane; (g) STrIPS membrane collection. **B** Holder assembly for the precursor deposition device. The precursor deposition device is clamped onto the white support holder while the substrate is moving underneath.

## S4. Three-phase contact angle measurements

The contact angle measurements for particle-free precursor mixtures on PET are performed in water with the setup shown in Fig. S5-A. As the density of the particle-free precursor mixture is lower than for water, the precursor droplets float in the continuous phase. Thus, the precursor is dispensed from a needle below the PET substrate.

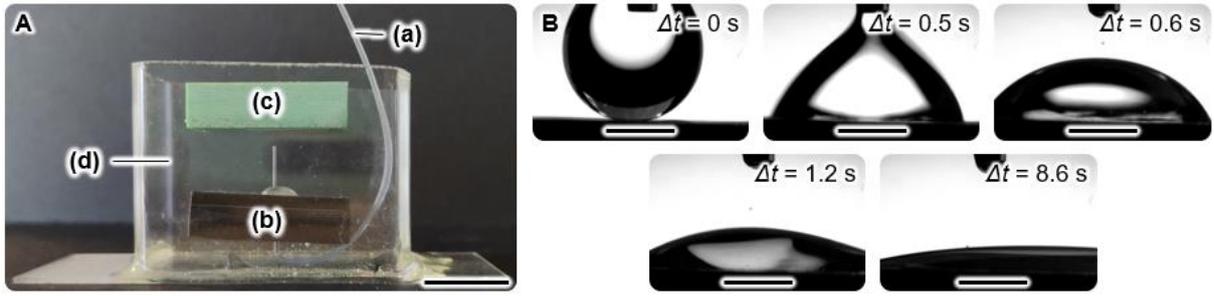

**Figure S5: A** Setup for measuring the three-phase contact angle of precursor mixtures on PET in water: (a) tubing to supply the precursor liquid; (b) dispensing needle; (c) holder with PET substrate at the bottom; (d) container filled with the continuous phase. Scale bar is 20 mm. **B** Spreading of a droplet of pure BDA on PET substrate in water (pH 3) over time. Scale bars are 2 mm.

### S5. STrIPS membrane thickness control

For the investigation of the STrIPS membrane thickness a homogeneous precursor mixture composed of the liquids BDA, water, and methanol (MeOH) with composition $\varphi_{BDA} = 0.40$, $\varphi_{Water} = 0.16$, and $\varphi_{MeOH} = 0.44$ is prepared containing 11 wt % nanoparticles and 15 mM $CTA^+$. The lower viscosity of the BDA/water/MeOH mixture compared to a BDA/water/EtOH precursor (Fig. S6) can potentially result in thinner STrIPS membranes due to enhanced outflow of the precursor out of the deposition device. However, both the precursor flow rate and the substrate pulling speed are the main factors that govern the amount of precursor which is deposited onto the substrate.

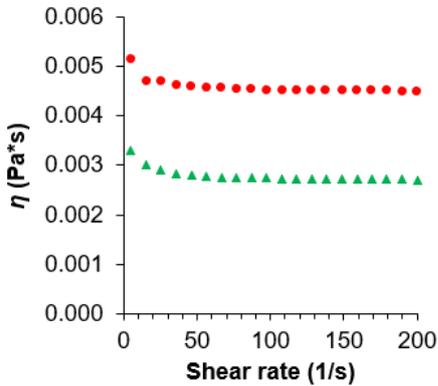

**Figure S6:** Dynamic viscosity $\eta$ (20 °C) for BDA/water/MeOH (▲) and BDA/water/EtOH (●) at different shear rates. For viscosity measurements a cone-plate compact rheometer (PHYSICA MCR 300, Anton Paar) with a cone angle of 1° is used.

The STrIPS membrane thickness can be calculated considering a finite volume element of the precursor mixture: The flow rate $Q_P$ (m³/s) at which the precursor flows through the deposition slit can be expressed by the cross-sectional area of the slit $A$ (m²) multiplied by the length of the volume element $y$ (m) over time $t$ (s; Eq. 1).

$$Q_P = \frac{A * y}{t} \qquad (Eq.\ 1)$$

The term $y / t$ is the velocity $v_P$ (m/s) at which the precursor flows through the deposition slit (Eq. 2).

$$Q_P = v_P * A \qquad (Eq.\ 2)$$

When the precursor exits through the deposition slit and contacts the substrate it is assumed that the velocity $v_P$ of the precursor mixture is equal to the substrate pulling speed $u_S$ (m/s; Eq. 3).

$$Q_P = u_S * A \qquad (Eq.\ 3)$$

The cross-sectional area of the membrane $A$ can be calculated using the width $w$ (8 mm) of the deposition slit and the thickness of the membrane $d$ (Eq. 4).

$$A = w * d \qquad (Eq.\ 4)$$

Thus, the thickness $d$ of the STrIPS membrane is given by Eq. 5 as:

$$d = \frac{A}{w} = \frac{Q}{u_S * w} \qquad (Eq.\ 5)$$

## S6. Fluorescence emission spectra of Rhodamine 110 chloride and Nile red

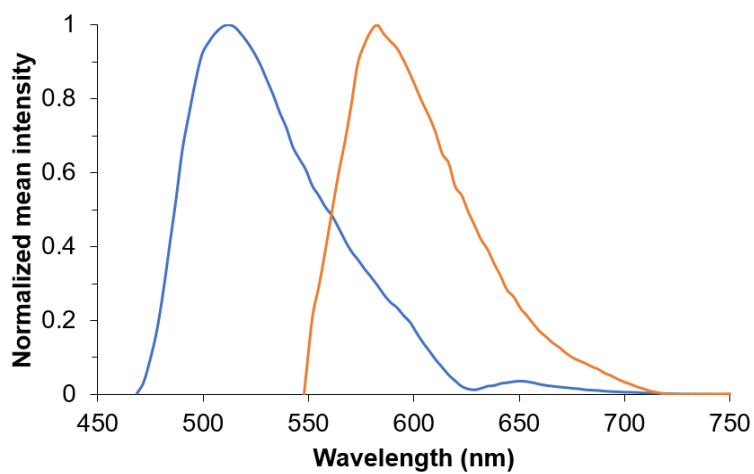

**Figure S7:** Normalized fluorescence emission spectra. In blue the fluorescence of Rhodamine 110 chloride is shown upon excitement with 488 nm laser light. The orange spectrum shows the fluorescence of Nile red excited with 561 nm laser light. The fluorescence emission spectra are acquired by local probing of a fluorescence dye-labelled STrIPS membrane immersed in diethyl phthalate.

## S7. STrIPS membrane structure evolution and transient solvent diffusion modelling

The structural evolution of the STrIPS membrane casted from precursor *iii* is tracked by arresting the BDA/water phase separation at different stages. To this end, the membrane is UV-polymerized at various travelling distances in the water bath. The UV-polymerization of the BDA arrests the membrane structure instantly [1]. The membrane travelling time $t$ (s) before polymerization is calculated from the precursor flow rate $Q_P$ (m³/s), the cross-sectional area of the membrane $A$ (m²) and the membrane travelling distance $s$ (m) which is determined by the position of the UV-lamp in the water bath (Eq. 6).

$$t = \frac{s * A}{Q_P} \qquad (Eq.\ 6)$$

This gives the correlation between travelling distance and travelling time of the membrane in Table S1.

**Table S1:** Membrane travelling distance before UV-polymerization and corresponding travelling time.

| $s$ (mm) | 30 | 60 | 120 | 240 |
|---|---|---|---|---|
| $t$ (s) | 0.6 | 1.2 | 2.4 | 4.8 |

To visualize the progress of the phase separation, the membrane cross-section is imaged via scanning electron microscopy (SEM). The membrane structure is compared with a solvent diffusion model to obtain the concentration profiles for ethanol during the STrIPS membrane formation (COMSOL® Multiphysics version 5.5). The membrane is modelled as 2D rectangular geometry with the Physics engine "Transport of Diluted Species" under consideration of the present flow geometry and velocities. Regarding the geometry, a membrane thickness of 50 µm is chosen based on the SEM analysis of the STrIPS membrane structure evolution (see below) and a 500 µm wide section is modelled as this represents the ethanol concentration profiles across the entire width of the membrane. The model couples flow and ethanol diffusion from the precursor into the surrounding water via the membrane surface. The ethanol is unable to diffuse through the PET substrate at the bottom of the membrane, thus, a no-flux boundary condition is selected for this part of the membrane. To account for the fact that the membrane is continuously moving through a large water reservoir, the ethanol concentration in the water bath is set to 0 vol %. The mesh showing the nodes for the numerical calculation of the ethanol concentration is depicted in Fig. S8-A. Convection of water is neglected because radial diffusion dominates over axial dispersion as shown in detail in the SI (p. 18-19) of reference [2].

The COMSOL® model uses Fick's second law of diffusion, describing the ethanol concentration $c_E$ in the precursor mixture with the differential equation Eq. 7:

$$\frac{\partial c_E}{\partial t} = D_E \frac{\partial^2 c_E}{\partial r^2} \qquad (Eq.\ 7)$$

At the beginning of the simulation (t = 0.0 s), the membrane interior has an ethanol concentration of $c_E^0 = \frac{\varphi_E^0 \cdot \rho_E}{M_E} = \frac{0.427 \cdot 790 \frac{kg}{m^3}}{0.046 \frac{kg}{mol}} = 7333 \frac{mol}{m^3}$ (corresponding to 43 vol %). The transient diffusion model is refined by choosing concentration dependent diffusion coefficients. The concentration dependence of the diffusion coefficient $D_E$ for ethanol in water is plotted in Fig. S8-B [3], [4]. For the diffusion simulation, $D_E$ is evaluated via linear interpolation between the data points in Fig. S8-B to solve Eq. 8.

$$\frac{\partial c_E}{\partial t} = D_E(c_E) \frac{\partial^2 c_E}{\partial r^2} \qquad (Eq.\ 8)$$

The partitioning coefficient for ethanol between the BDA- and the water-rich phase is set to 1 [1]. Furthermore, STrIPS membranes typically form a surface with a lowered porosity compared to the interior [1]. This is incorporated in the model as thin diffusion barrier on the membrane surface. We therefore assume over a thickness of 3 µm below the membrane surface a diffusion coefficient of $D_E' = D_E(c_E)/10$ [1], [5].

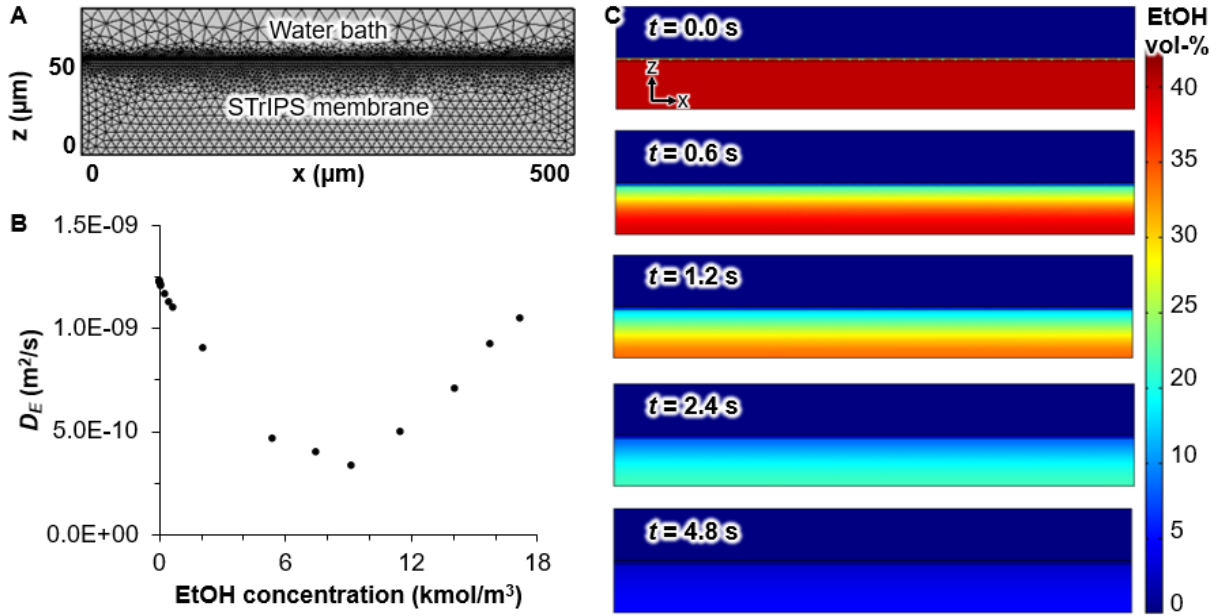

**Figure S8: A** Mesh representing nodes for the numerical calculation of the ethanol (EtOH) concentrations. **B** Diffusion coefficients $D_E$ for EtOH in water in dependence of the EtOH concentration $c_E$. **C** Concentrations plots of EtOH at the membrane cross-section at different time steps $t$ of the simulation.

The simulation results in Fig. S8-C show that within the first 0.6 s of the STrIPS process the ethanol concentration drops primarily in the uppermost part of the membrane. We find that at 10 µm from the membrane surface, the concentration is lowered from initially 43 vol % to ~23 vol %. At 1.2 s, the ethanol concentration has also started to decrease in the lower regions of the membrane. In the following seconds, the ethanol continues to diffuse out from the bottom of the membrane as observed in the concentration plots for 2.4 s and 4.8 s. It takes around 5 s for the ethanol to fully diffuse out of a 50 µm thick membrane.

The gradual diffusion of ethanol from the membrane interior to the surrounding water is reflected in the progress of the BDA/water phase separation. Fig. S9 shows SEM images of the membrane cross-sections acquired at different time steps of the membrane formation which thus correspond to different stages of the STrIPS process. The phase separation grows 11 µm below the membrane surface within the first 0.6 s, and after 1.2 s the phase-separated region stretches 14 µm below the surface. After 4.8 s the membrane structure formation is completed across the membrane depth of 50 µm.

Our model is in basic agreement with the experimental investigation of the structure formation viewing that the phase-separated region is initially only found at the membrane top surface and extends deeper into the membrane over time. However, the model predicts a low ethanol concentration at the bottom of the membrane already after 2.4 s, but Fig. S9 shows that the membrane's lower half is still unstructured and therefore has not yet phase separated. This discrepancy potentially results from the fast stabilization of a porous structure at the top of the membrane during STrIPS. This dense structure hinders the solvent from diffusing out of the membrane and delays the phase separation.

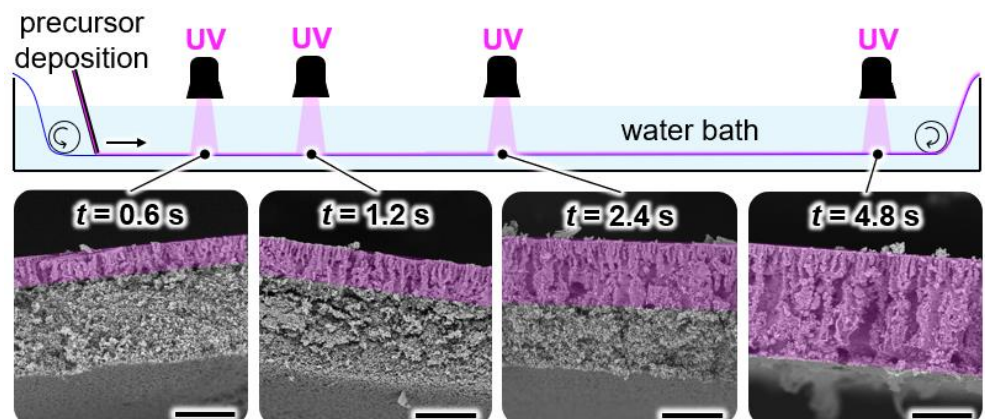

**Figure S9:** Structure evolution of the STrIPS membrane. SEM micrographs of the membrane cross-section are shown for different travelling times $t$ of the membrane before UV-polymerization in the water bath. All scale bars are 25 µm.

**S8. Pore size analysis**

The pore sizes of the STrIPS membrane are measured by confocal image analysis using the software Fiji ImageJ (version 1.53k14) with the processing workflow depicted in Fig. S10. The confocal micrographs presented in Fig. 5B in the manuscript are imported into the software. First, the contrast between the polyBDA and the water domains is enhanced and the confocal micrograph is converted to an 8-bit image. After applying a bandpass filter and thresholding of the image to highlight the water pores, the image is binarized such that the water pores are shown as white pixels against a black background. To measure the area of the pores, the command "Analyze Particles" is selected. For the confocal micrograph of the membrane for 14 wt % nanoparticles ($z = 50$ µm; Fig. 5B) small nucleation droplets $< 1$ µm$^2$ in size are left out of the measurement by setting a lower size limit for this analysis step. The analysis produces an outline of the detected pores and a table listing the size of every pore. To confirm that the membrane pores have correctly been identified by the software, the outlines of the pores are placed over the original confocal micrograph. Based on this procedure, the average pore size is calculated.

The pore size analysis was not performed for confocal micrographs from the bottom of the membrane ($z = 95$ µm). At these membrane depths, the fluorescence of polyBDA becomes too weak to unambiguously distinguish the water pores from the polyBDA domains and thus to accurately determine the membrane pore size.

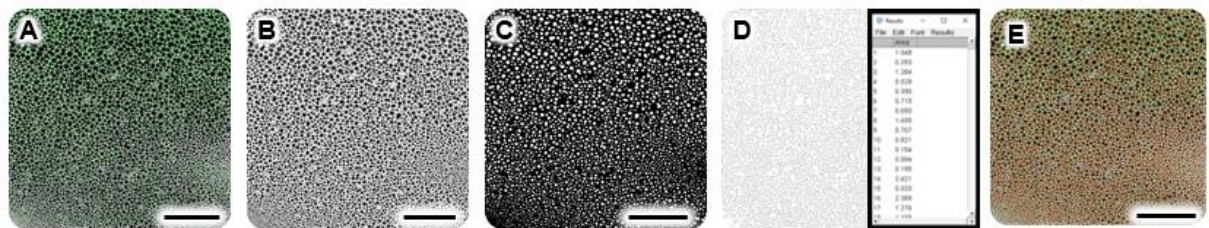

**Figure S10: A** Importing the confocal micrograph into ImageJ. Water is shown in black, polyBDA in magenta and nanoparticles in green. **B** Enhancing the contrast between polyBDA and water and converting the micrograph into an 8-bit image. **C** Binarization of the image after running a bandpass filter and pixel thresholding. The water pores are white, the polyBDA domains black. **D** Automated pore identification and pore size measurement. **E** The outlines of the identified pores are placed over the original confocal micrograph to assess the quality of the analysis. All scale bars are 25 µm.

**Supplementary References**